\documentclass{emulateapj}

\usepackage{epsfig}

\newcommand\rmc{{\rm c}} 
\newcommand\rms{{\rm s}} 
\def\rmp{{\rm p}}
 
\newcommand\rmO{{\rm O}} 
\newcommand\rmS{{\rm S}}
\newcommand\rmP{{\rm P}} 
\newcommand\rmD{{\rm D}} 
\newcommand\RR{{\rm RR}}

\shortauthors{N. R. Badnell}
\shorttitle{RR data}

\begin{document}

\title{Radiative recombination data for modelling dynamic finite-density plasmas}

\author{\sc  N. R. Badnell}

\affil{Department of Physics, University of Strathclyde, Glasgow G4 0NG, UK}

\begin{abstract}
We have calculated partial final-state resolved radiative recombination (RR) rate coefficients
from the initial ground and metastable levels of all elements up to and including  Zn, plus Kr, Mo,
and Xe, for all isoelectronic sequences up to Na-like forming Mg-like. The data are archived according
to the Atomic Data and Analysis Structure (ADAS) data class {\it adf48}, which spans a 
temperature range of $z^2(10^1 - 10^7)$~K, where $z$ is the initial ion charge. Fits to total
rate coefficients have been determined, for both the ground and metastable levels, and
those for the ground are presented here. Comparison is made both with previous RR rate 
coefficients and with (background) $R$-matrix photoionization cross sections.
This RR database complements a DR database already produced and both are being used to produce
updated ionization balances for both (electron) collisionally ionized and photoionized plasmas.
\end{abstract}


\section{Introduction}
\label{sec:Intro}
Astrophysical and laboratory plasmas which are not in (local) thermodynamic equilibrium
require a detailed modelling of their atomic and, sometimes, molecular reactions so as to
determine the level populations and ionization balance of their constituents. These in turn
are basic ingredients for the physical and spectral diagnostic modelling of stellar coronae,
gaseous nebulae, supernova remnants, fusion plasmas, etc. At low particle densities, e.g. the `coronal approximation',
ionization balance is between total groundstate dielectronic-plus-radiative recombination and
groundstate ionization by electrons and/or photons, while level populations are determined by collisional
excitation from the groundstate and radiation to all possible lower states. As the particle
density increases, a separation between the two is no longer possible and level populations,
effective recombination \& ionization rate coefficients and ionization balance are
determined by solving the collisional--radiative population rate equations \citep{BKM62, BS69, Sume06}.
Furthermore, even at the low densities found in gaseous nebulae, for example, the populations of ions of 
C, N, O which have fine-structure levels in their ground term are not concentrated in the ground level.
Recombination from excited levels within the ground term is quite different (smaller) compared
to that for the ground level, due to the presence of additional autoionization pathways.
Furthermore, transient plasmas such as solar flares do not have time to establish quasi-static equilibrium with
the ground level and so require metastable levels (and the excited-state populations built upon them) to be
treated on an equal footing as the ground level when it comes to collisional--radiative modelling \citep{Sum83}.

However, historically, total (zero-density) ground state recombination rate coefficients have dominated 
the literature --- see, for example, those used by \citet{Shul82a, Arna85a, Peq91, Arna92a, Mazz98a}. A systematic
attempt to move beyond this picture has been described by \citet{Badn03a}, who focussed on dielectronic
recombination but much of the discussion there is applicable to radiative recombination --- deliberately so.
Now, using {\sc autostructure} \citep{Badn86a}, we have calculated partial final-state resolved 
radiative recombination  (RR) rate coefficients from the initial ground 
and metastable levels of all elements up to and including Zn, plus Kr, Mo and Xe, for
all isoelectronic sequences up to Na-like forming Mg-like. The data are archived according
to the Atomic Data and Analysis Structure (ADAS) data class {\it adf48}\footnote{The historic ADAS 
dataclass file for RR is {\it adf08} but it is  structured to handle RR to low-lying levels only, 
so as to supplement their population by electron-impact excitation.} \citep{Sum05}. Fits to total RR
rate coefficients have been determined, for both the ground and metastable levels, and
those for the ground are presented here. All of the data are available online \citep{Badn06b}, along with
corresponding data for DR \citep{Badn06a}.

Previous work on RR has been summarized by \citet{Mazz98a}. Since then, \citet{Gu03b} has presented RR
rate coefficients obtained using his Flexible Atomic Code {\sc fac} for Mg, Si, S, Ar, Ca, Fe, and Ni,
for all isoelectronic sequences through to F-like forming Ne-like. 

The time reversed process of RR (plus DR) is photoionization (PI),
indeed, our RR is determined from our direct photoionization cross sections on using the principle
of detailed balance. We make comparison
of our groundstate-to-groundstate photoionization cross sections of neutral atoms with those obtained
from various $R$-matrix calculations, by others, as the most severe
test of our results. Similar $R$-matrix data has been used by Nahar and co-workers \citep[e.g.,][]{Nah99}
to form total (DR+RR) rate coefficients, and we make a comparison with it by combining our RR data with 
complementary DR data \citep{Badn06a}.

The remainder of the paper is organized as follows: in Section 2 we describe our methodology, 
in Section 3 we present our results and make comparisons with the results of earlier works, and then
make some concluding remarks.

\section{Methodology}
\subsection{Theory}
We use the independent processes approximation so as to treat RR and DR separately. Interference
between DR resonances and the RR background can safely be neglected, especially on integrating
over resonance profiles \citep{Pind92a}.
We use {\sc autostructure} \citep{Badn86a} to calculate multi-configuration intermediate-coupling
distorted-wave photoionization cross sections \citep{Badn03b}. These are converted to RR 
cross sections using detailed balance. For elements beyond Zn we use
semi-relativistic radial functions in place of the non-relativistic ones used up until then \citep{Pind90}. 

\subsection{Atomic Details}
For each sequence, we used the same set of initial and final configurations as was done for ($\Delta n=0$) DR 
\citep{Badn03a}, i.e., we included configuration mixing in the initial $N$-electron, and final $N$- or 
$(N+1)$-electron complexes for RR outwith or into the initial complexes, respectively.
Each inequivalent $nl$ electron-state was treated separately. Radial functions were determined using the Slater-Type-Orbital
model potential option of {\sc autostructure}, which generates (slightly, in this case) non-orthogonal
orbitals. We follow \citet{Cow81} in not imposing orthogonality and assuming the overlaps to be zero or unity,
as appropriate. This is more of an issue for inner-shell photoionization where large relaxation effects
occur for L-shell orbitals following K-shell photoionization, see e.g., \citet{Gor06}.
For recombination/ionization to/from  the ground complex
we used the observed ionization potentials from \citet{NIST}. Energy adjustments were made to the diagonal
of the $LS$-coupling Hamiltonian before recoupling to, and diagonalization of, the intermediate coupling
Hamiltonian. Finally, to determine total RR and/or DR rate coefficients we sum over only final states which are
stable against autoionization.

\subsection{Numerics}
Cross sections were calculated at zero electron energy and then on
a $z$-scaled logarithmic energy mesh with 3-points per decade spanning $(2.1\times 10^{-6} -
1\times 10^{2})z^2$ Ryd, where $z$ is the charge of the ion before/after recombination/photoionization. 
Tests were also made using 5-points per decade. The length gauge is used at low energies, progressively
and automatically switching to velocity and then acceleration as the energy increases so as to
maintain numerical accuracy, but not before the succeeding gauge has converged to the preceding one. 
Cross sections at
higher energies were determined by extrapolation as $E^{-l-7/2}$ (for PI), where $l$ is the orbital angular
momentum of the bound orbital. Rate coefficients were determined over $(10^1 - 10^7)z^2$~K and so are
not particularly sensitive to the exact form used for the extrapolation of the cross sections. 
With many thousands of individual partial rate coefficients to
be determined, a fast and accurate Mawellian convolution method is desirable. We use a variant of that due 
to \citet{Bur64},
viz., a 5-point Newton-Coates quadrature on a doubling energy mesh, with a 4-point Lagrange interpolation
formula to convert from the initial logarithmic energy mesh, completed by an exponential integral.
By using a $z$-scaled minimum temperature we can use a fixed maximum value for the principal
quantum number, viz., 1000, and orbital angular momentum, viz., 200. We use analytic hydrogenic radial integrals
for $l\ge 4$ making, use of the fast, accurate, recurrence relations of \citet{Bur64}. All resultant
partial RR rate coefficients are designed to be numerically accurate to 3 significant figures.

\subsection{Database Details}
The processing of photoionization cross  sections from {\sc autostructure} is carried-out using the {\sc adasrr}
code.  The output of {\sc adasrr} is similar in nature to the {\sc adasdr} code used for DR \citep{Badn03a}, i.e. 
the ADAS {\it adf48} file
which it writes for RR is very similar in structure to the {\it adf09} file for DR \citep{Sum05}. Viz.,
for each initial level (ground-plus-metastables) there is a progressive bundling over final outer quantum 
numbers as $n$ increases. Partial RR rate coefficients to low-lying levels (those with $n\le 8$) are fully $J$-resolved,
then bundled-$nl$ ($n\le 10$) and bundled-$n$ ($n<1000$), but each with a fully $J_\rmc$-resolved core.
Finally, total RR rate coefficients are written for each specified initial ground and metastable level.
The full set of RR data --- partials ({\it adf48} files) and totals (fit coefficients) --- is available online
\citep{Badn06b}, alongside the DR data \citep{Badn06a}, and it is also part of the ADAS distribution. (OPEN\_ADAS,
a public access version of ADAS is scheduled to come online in the Summer of 2006, sponsored and hosted by the
International Atomic Energy Agency.)

\subsection{The Fit}
Total RR rate coefficients, for both ground and metastable initial levels, are fitted to the
usual functional form \citep{Ver96, Gu03b} viz.
\begin{eqnarray}
\label{fit1}
&\alpha&_\RR(T)=A  \\
&\times& \left[\sqrt{T/T_0}\left(1+\sqrt{T/T_0}\right)^{1-B}\left(1+\sqrt{T/T_1}\right)^{1+B}\right]^{-1},
\nonumber
\end{eqnarray}
where, for low-charge ions, $B$ is replaced as
\begin{equation}
\label{fit2}
B \rightarrow B + C \exp(-T_2/T ).
\end{equation}
Here, $T_{0,1,2}$ are in units of temperature (K) and the rate coefficient $\alpha_\RR(T)$ is in units of cm$^3$~s$^{-1}$.
(The units of $A$ are also cm$^3$~s$^{-1}$, while $B$ and $C$ are dimensionless.) 
A non-linear least-squares fit was used to determine the coefficients ($A, B, T_{0,1}; C, T_2$). The fits
are accurate to better than 1\% over $(10^1 - 10^7)z^2$~K for multiply charged ions,
5\% for singly and doubly ionized, and have the correct asymptotic forms outside of this range.

\section{Results}

Fit coefficients according to Equations (\ref{fit1}) and (\ref{fit2}) are presented in Table \ref{tab1} for
total groundstate RR rate coefficients for all elements up to and including Zn, plus Kr, Mo and Xe, for
all isoelectronic sequences up to Na-like forming Mg-like. 
Partial final-state resolved RR rate coefficients for initial ground and metastable levels
are archived as ADAS {\it adf48} files \citep{Sum05}.  Fits to total RR
rate coefficients have been determined for initial metastable levels as well.
All of the data are available online \citep{Badn06b}, along with corresponding data for DR \citep{Badn06a}.

\subsection{Photoionization Cross Sections}
The description of groundstate-to-groundstate photoionization of neutral atoms provides the most
stringent test of the accuracy of our description of partial and high-temperature-total  RR
rate coefficients. 

\begin{figure}
  \begin{center}
  \epsfig{file=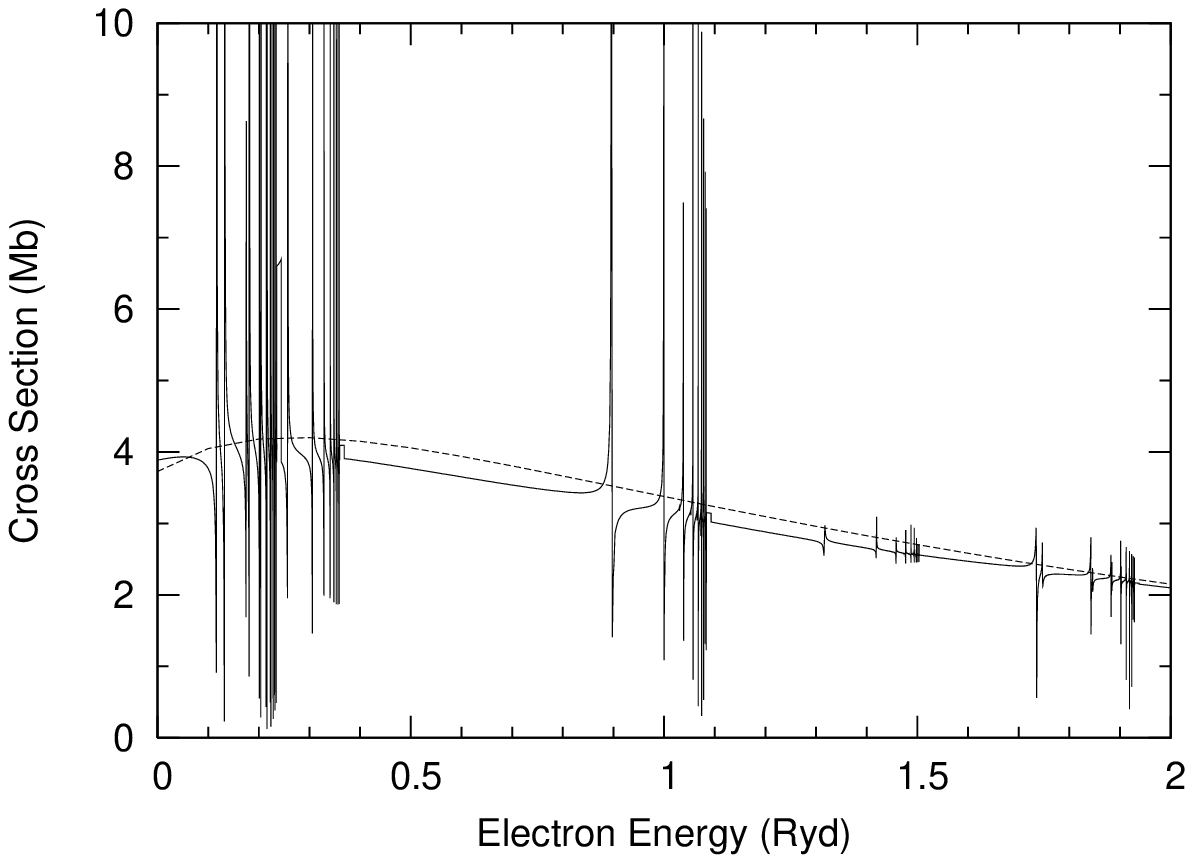}
  \end{center}
  \caption[]{Photoionization: $\rmO(^3\rmP)\rightarrow \rmO^+(^4\rmS)$. Solid curve, $R$-matrix
  results \citep{But90}; dashed curve, present {\sc autostructure} results.}
  \label{Fig_O}
\end{figure}

In Figure~\ref{Fig_O} we compare photoionization cross sections of neutral O for 
the transition $\rmO(^3\rmP)\rightarrow \rmO^+(^4\rmS)$. We see that our {\sc autostructure} results closely track
the background $R$-matrix cross sections of \citet{But90}, to within 7\%.

\begin{figure}
  \begin{center}
  \epsfig{file=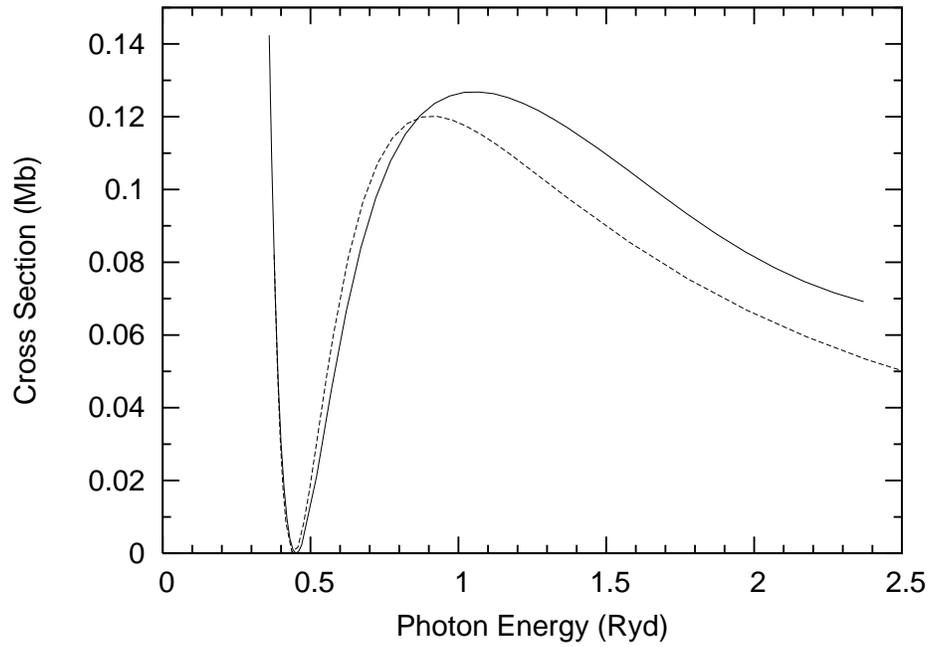}
  \end{center}
  \caption[]{Photoionization of Na (3s). Solid curve, $R$-matrix
  results \citep{TOP}; dashed curve, present {\sc autostructure} results.}
  \label{Fig_Na}
\end{figure}

In Figure~\ref{Fig_Na} we make a similar comparison for neutral Na. This is an interesting system
since it features a Cooper minimum\citep{Fan68}. We see that our {\sc autostructure} photoionization 
cross sections give a reasonable description of this feature, on comparison with the $R$-matrix results of
\citet{But83,TOP}.

\begin{figure}
  \begin{center}
  \epsfig{file=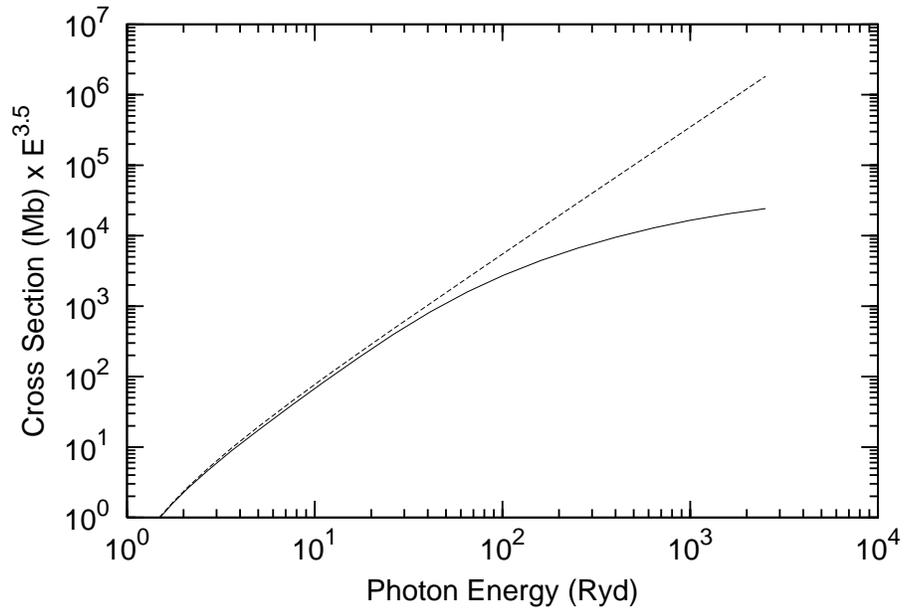}
  \end{center}
  \caption[]{Scaled photoionization cross section of Mg$^+$(3s).
  Solid curve, present {\sc autostructure} results; dashed curve, Dirac--Hartree--Slater
  results of \citet{Ver95}.}
  \label{Fig_Mgh}
\end{figure}

Finally, in Figure~\ref{Fig_Mgh} we look at the high energy behaviour of the Mg$^+$ photoionization
cross section: we compare our energy-scaled {\sc autostructure} cross sections with the 
Dirac--Hartree--Slater results  of \citet{Ver95}. We see
that at the high energy cross sections of \citet{Ver95} have still to reach their asymptotic form,
of $E^{-7/2}$. Analysis of the fitting form and parameters of \citet{Ver95} shows that it will not
reach asymptopia until $\sim 10^6$ Ryd, which appears unphysically high. Similar disagreement
is seen in other low-charge Na-like ions, the effect decreasing with increasing residual charge.

\subsection{Total Groundstate Rate Coefficients}

\begin{figure}
  \begin{center}
  \epsfig{file=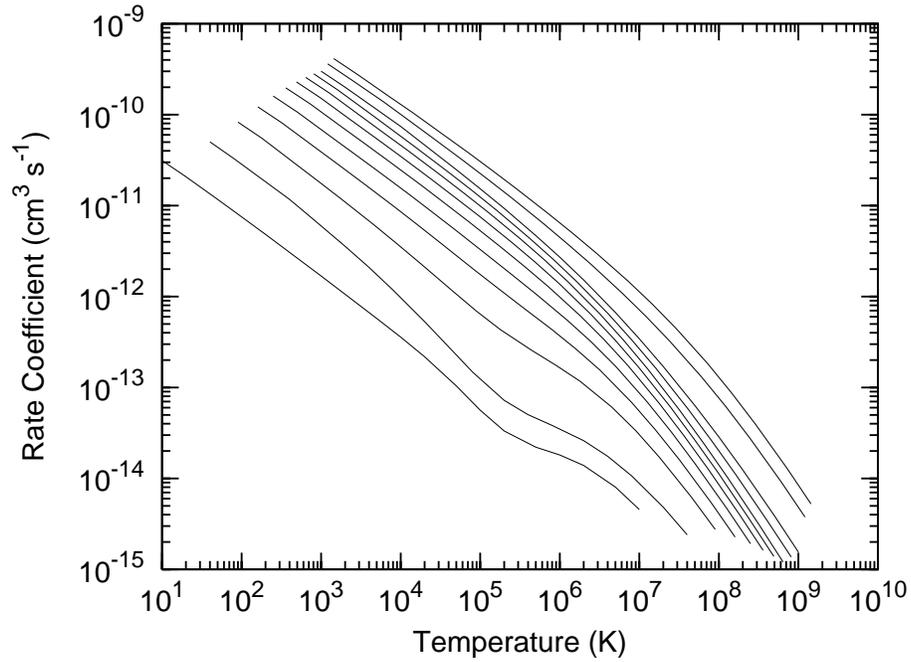}
  \end{center}
  \caption[]{Total groundstate RR rate coefficients for the Mg isonuclear sequence (present
  {\sc autostructure} results). Curves for singly ionized to bare are shown, moving left-to-right.}
  \label{Fig_Mg}
\end{figure}

\begin{figure}
  \begin{center}
  \epsfig{file=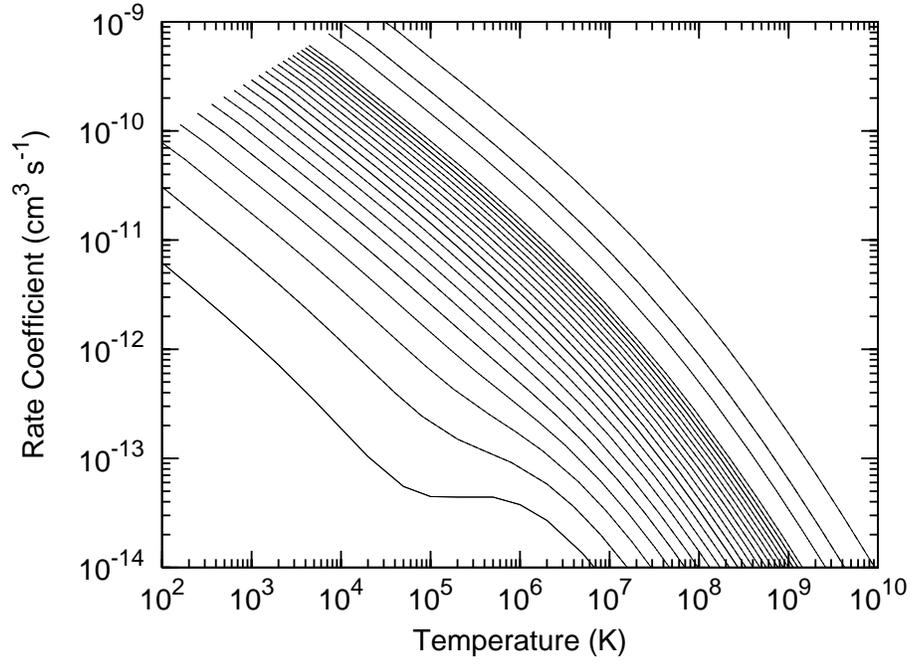}
  \end{center}
  \caption[]{Total groundstate RR rate coefficients for the F isoelectronic sequence (present
  {\sc autostructure} results). Curves for Ne to Zn, plus Kr, Mo, Xe are shown, moving left-to-right.}
  \label{Fig_flike}
\end{figure}

In Figures~\ref{Fig_Mg} and \ref{Fig_flike} we illustrate an overview of total groundstate
RR rate coefficients, viz. the Mg isonuclear sequence and the F isoelectronic sequence. The
change in behaviour on moving from K- to L- and then M-shells is clearly seen in the former
and the characteristic high-temperature bump at low-charges in the latter.

\begin{figure}
  \begin{center}
  \epsfig{file=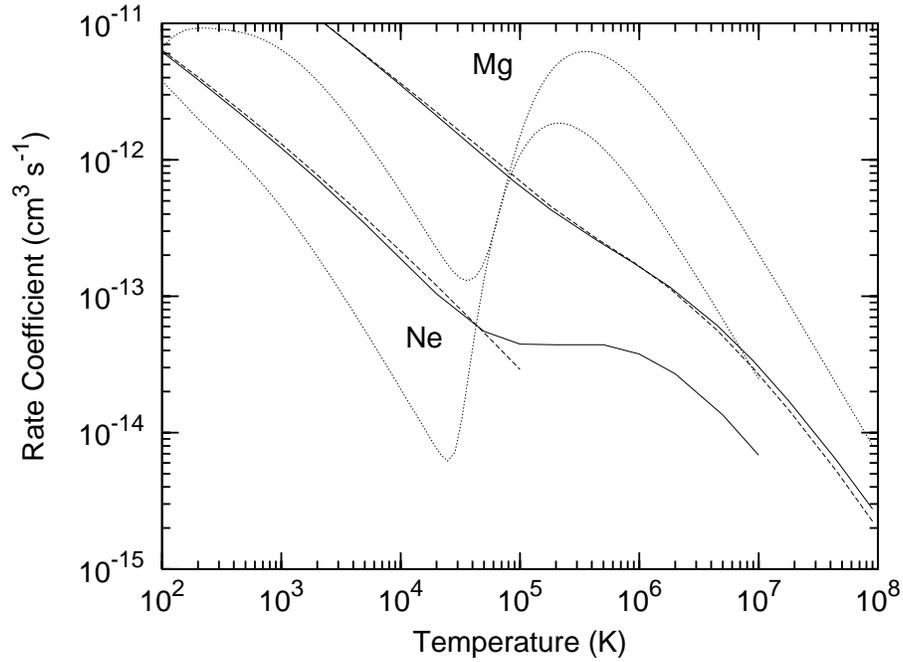}
  \end{center}
  \caption[]{Total groundstate rate coefficients for F-like Ne and Mg. Solid curves,
  present {\sc autostructure} RR results; dashed curves, RR results of \citet{Peq91} and 
  \citet{Gu03b} for Ne and Mg, respectively; dotted curves, DR results of \citet{Zat06}.}
  \label{Fig_NeMg}
\end{figure}

In Figure~\ref{Fig_NeMg} we compare our F-like Ne and Mg rate coefficients from {\sc autostructure}
with those of
\citet{Peq91} and \citet{Gu03b}, respectively.  The high-temperature limit on the validity of the
simple functional form used by \citet{Peq91} is a few times $10^4$~K, but DR dominates thereon. The more general
form, given by Equations (\ref{fit1}) and (\ref{fit2}), does not suffer from this limitation.
The comparison with the results of \citet{Gu03b} for Mg illustrates the worst agreement
found with his work, viz., about 10\% in the vicinity of $5\times 10^4$~K, and up to 20\% at $10^8$~K.

\begin{figure}
  \begin{center}
  \epsfig{file=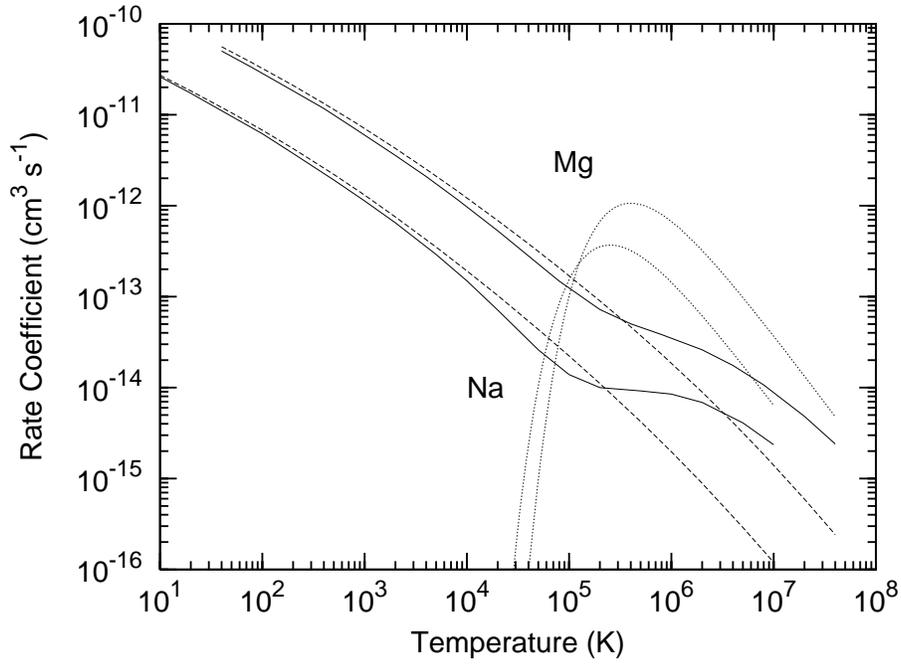}
  \end{center}
  \caption[]{Total groundstate rate coefficients for Ne-like Na and Mg. Solid curves,
  present {\sc autostructure} RR results; 
  dashed curves, RR results of \citet{Ver96}; dotted curves, DR results of \citet{Gor06b}
  .}
  \label{Fig_NaMg}
\end{figure}

In Figure~\ref{Fig_NaMg} we compare our Ne-like Na and Mg rate coefficients from {\sc autostructure}
with those of \citet{Ver96}. 
The latter are based-on Opacity Project \citep{TOP} $R$-matrix background photoionization cross sections at low energies
and the Dirac--Hartree--Slater ones of \citet{Ver95} at higher energies. We have already seen the agreement between
our low energy (groundstate) neutral Na photoionization cross sections and those from $R$-matrix --- those for Mg$^+$ 
show good agreement as well (not shown). We noted also that the high energy Dirac--Hartree--Slater results lie {\it above}
those from {\sc autostructure}. But, here, we see that the high temperature RR rate coefficients of \citet{Ver96}
lie {\it below} our results --- RR into the 3s dominates at high temperatures --- which is puzzling.
However, the rise of DR at $10^5$~K  postpones significant disagreement in the (DR+RR) total until $\sim 10^7$~K.

\begin{figure}
  \begin{center}
  \epsfig{file=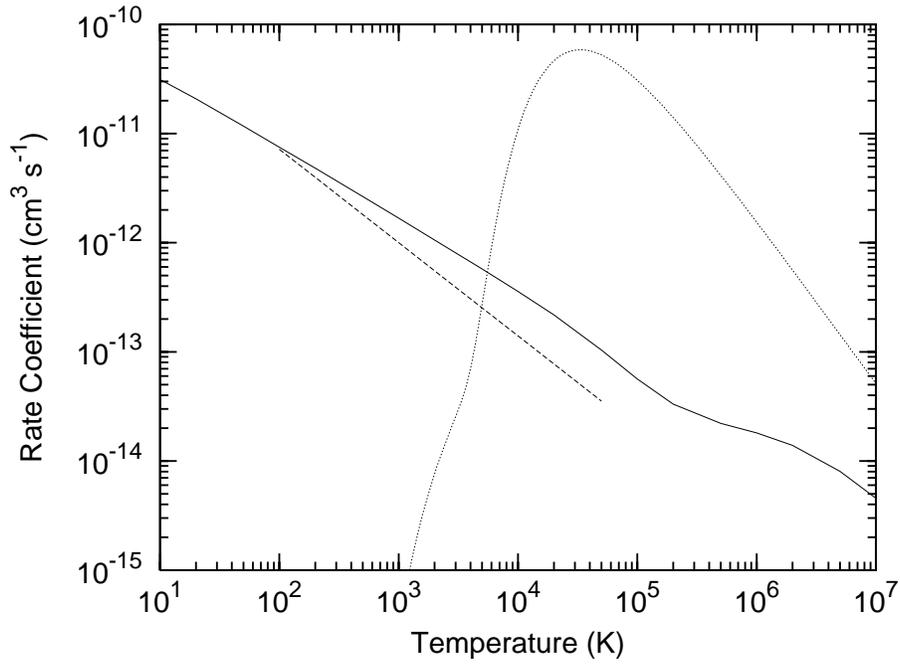}
  \end{center}
  \caption[]{Total groundstate rate coefficients for Mg$^{+}$. Solid curves,
  present {\sc autostructure} RR results; dashed curve, RR results of \cite{Ald73};
  dotted curves, DR results of \citet{Alt06}.}
  \label{Fig_Mg_Nalike}
\end{figure}

In Figure~\ref{Fig_Mg_Nalike} we compare our Mg$^+$ rate coefficients from {\sc autostructure}
with the RR of \citet{Ald73} \citep[and used again by][without modification]{Shul82a} and which still form
the basis of recommended data for this ion \citep{Mazz98a}. Our RR results are 70\% and a factor of 3
larger at 10$^3$ and 10$^4$~K, respectively, but DR dominates at the latter temperature while it is negligible 
at the former. The results of \citet{Ald73} make use of very restricted
(near-threshold) photoionization cross section measurements by \citet{Dit53}. This new rate coefficient changes
the ionization balance of Mg in many photoionized plasmas, and affects Mg II emissivities
noticeably \citep{Fer06}, for example.

\begin{figure}
  \begin{center}
  \epsfig{file=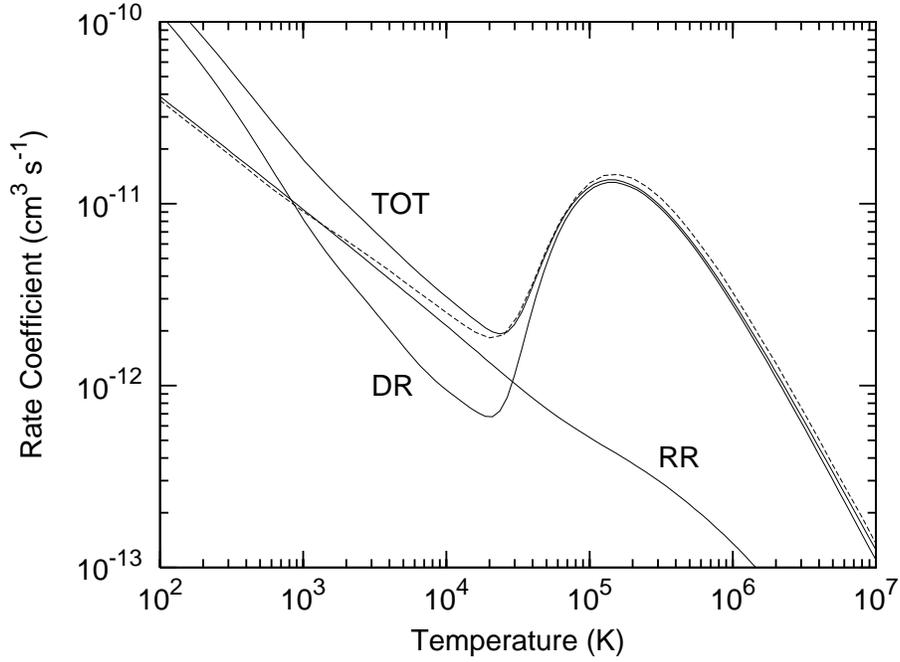}
  \end{center}
  \caption[]{Total groundstate rate coefficients for O$^{2+}$. Solid curves,  RR (present), DR
  \citep{Zat04a}, and
  total (DR+RR) from {\sc autostructure}; dashed curve, total from $R$-matrix \citep{Nah99}.}
  \label{Fig_O2N}
\end{figure}

Finally, in Figure~\ref{Fig_O2N} we compare total (DR+RR) rate coefficients from {\sc autostructure}
for O$^{2+}$ with the $R$-matrix results of \citet{Nah99}. We also show separately the contribution from the 
present RR rate coefficients and the DR ones of \citet{Zat04a}, taken from \citet{Badn06a}. 
At the high temperature peak ($2\times 10^5$~K) the results of \citet{Nah99} lie about 8\% above the
present ones, and remain constantly so thereon.
Below about $2\times 10^4$~K the contribution from fine-structure DR separates the two results. The
calculations of \citet{Nah99} were carried-out in $LS$-coupling and so this is absent and her results
track our RR results to within a few percent.

\subsection{Total Metastable Rate Coefficients}
\begin{figure}
  \begin{center}
  \epsfig{file=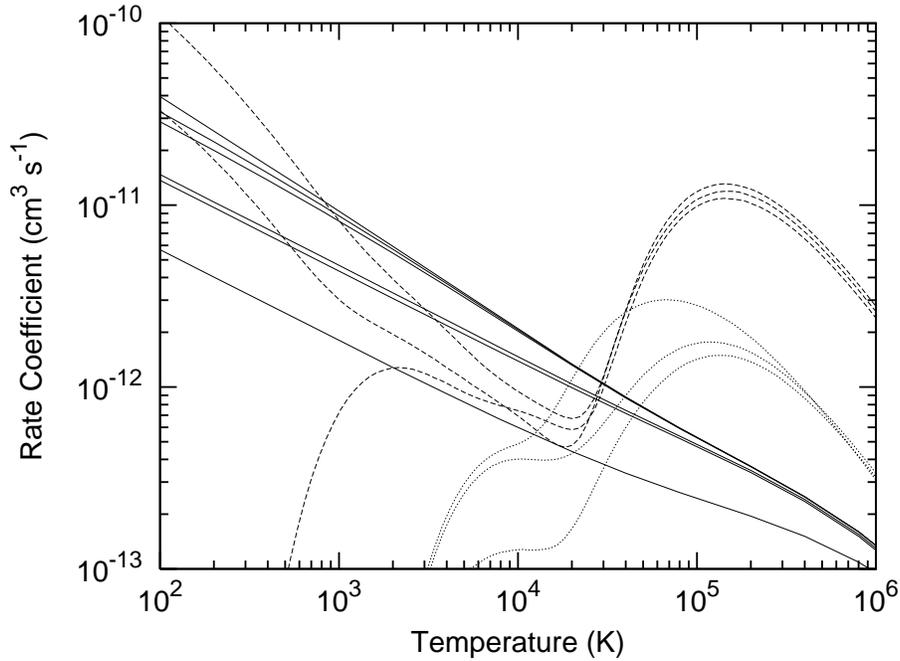}
  \end{center}
  \caption[]{Total rate coefficients from the ground and metastable levels of
O$^{2+} (^3\rmP_{0,1,2}, ^1\rmD_{2}, ^1\rmS_{0}, ^5\rmS_2 )$. Solid curves, RR,
top-to-bottom at $10^5$~K corresponds to increasing initial energy metastable; dashed curves, DR from
levels of the ground term, top-to-bottom corresponds to increasing initial energy metastable;
dotted curves, DR from the excited terms, bottom-to-top at $10^5$~K corresponds to increasing initial 
energy metastable. All DR taken from \citet{Badn06a}, based-on  \citet{Zat04a}.}
  \label{Fig_O2}
\end{figure}

In Figure~\ref{Fig_O2} we present total rate coefficients from the ground and metastable levels of
O$^{2+} (^3\rmP_{0,1,2}, ^1\rmD_{2}, ^1\rmS_{0}, ^5\rmS_2 )$. Fine structure splitting in the $^3\rmP_J$
ground term only starts to affect RR at very low temperatures ($\sim 30\%$ at 100~K). This is in contrast to DR
where the lower two levels are strongly enhanced by the fine-structure pathway(s) open to them while the
upper level only has a small low temperature peak which never exceeds the RR contribution. RR onto excited terms is 
reduced distinctly by autoionization, compared to that for levels within a term. The $^1\rmD_{2}$ and $ ^1\rmS_{0}$ 
contributions are similarly sized, both for RR and the high temperature DR peak (at $10^5$~K). The $^5\rmS_2$
RR contribution is the smallest but its high temperature DR peak exceeds that of the $^1\rmD_{2}$ and $ ^1\rmS_{0}$ 
initial states. The widely varying behaviours of DR and RR for ground and metastable levels is then reflected 
similarly in the temperature dependence of the (DR+RR) total recombination rate coefficients (not shown).

\subsection{Partial Rate Coefficients}
\begin{figure}
  \begin{center}
  \epsfig{file=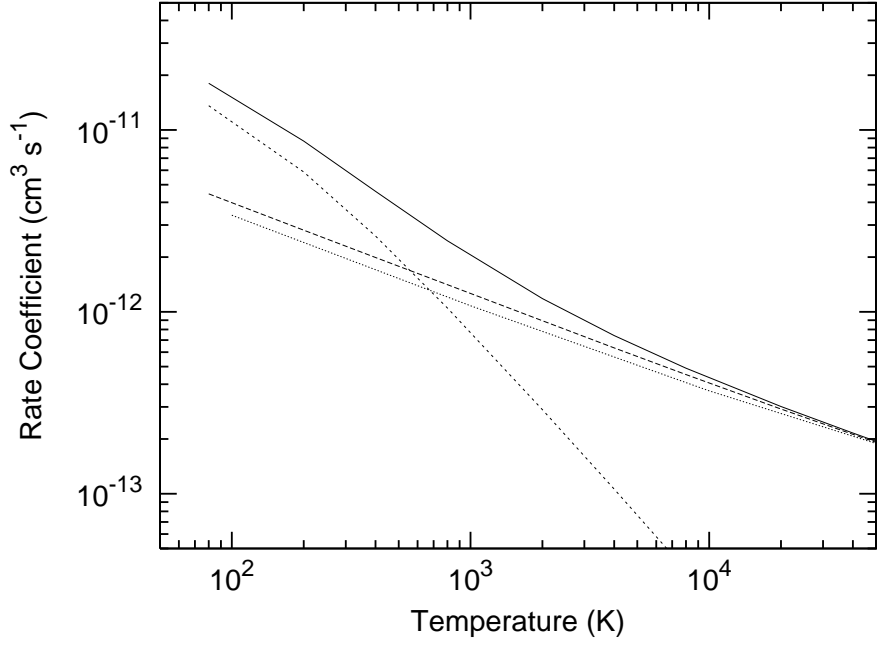}
  \end{center}
  \caption[]{Partial recombination rate coefficients for O$^{2+}$ to O$^+$ 
   groundstate-to-groundstate. Long-dashed curve, RR (present); short-dashed curve,
  DR \citep{Zat04a}; solid curve, total (DR+RR); dotted curve, $R$-matrix \citep{Nah99}.}
  \label{Fig_O2p1}
\end{figure}

\begin{figure}
  \begin{center}
  \epsfig{file=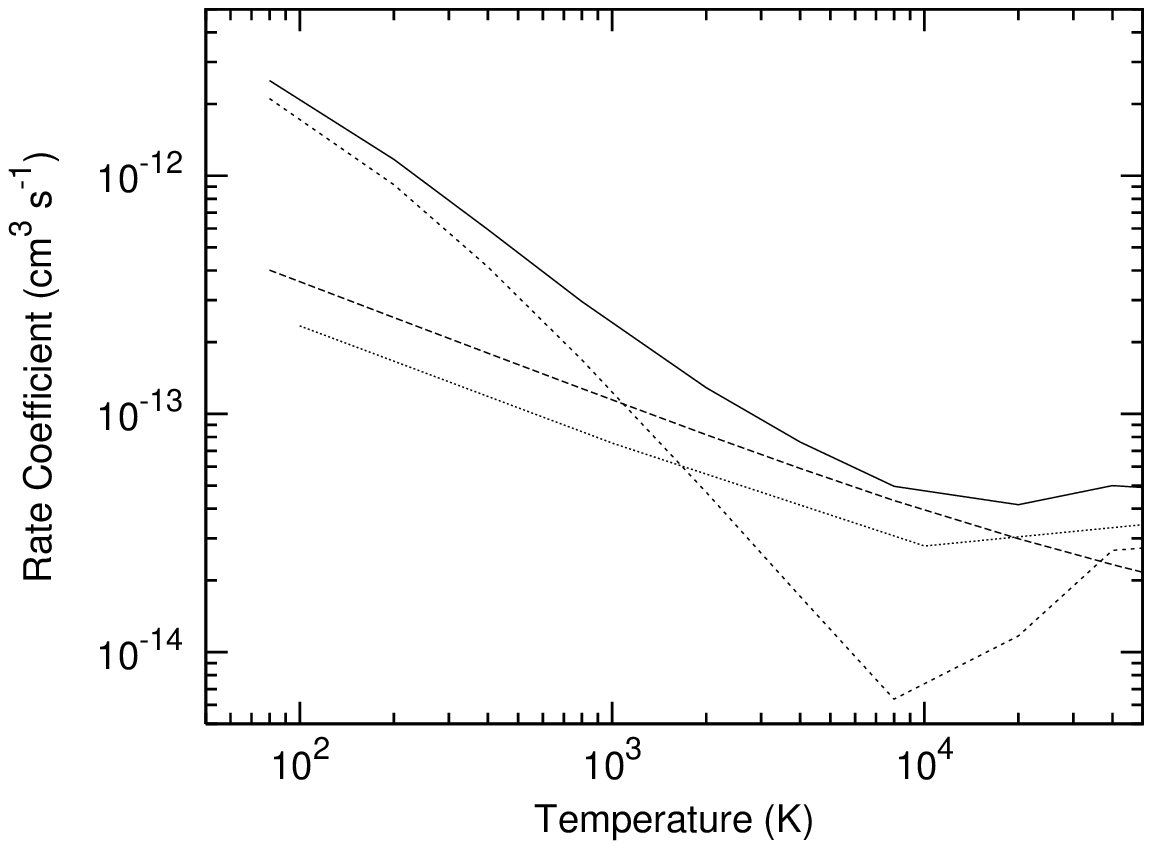}
  \end{center}
  \caption[]{Partial recombination rate coefficients for O$^{2+}$ groundstate to O$^+$ 
   $2\rms^22\rmp^23\rmp ^4\rmD$. Long-dashed curve, RR (present); short-dashed curve,
  DR \citep{Zat04a}; solid curve, total (DR+RR); dotted curve, $R$-matrix \citep{Nah99}.}
  \label{Fig_O2p4}
\end{figure}

In Figures~\ref{Fig_O2p1} and \ref{Fig_O2p4} we compare partial final-state resolved rate coefficients
for recombination from the initial ground state of O$^{2+}$ to the $2\rms^2 2\rmp^{3}\, ^{4}\rmS_{3/2}$
and $2\rms^22\rmp^23\rmp ^4\rmD_J$ states of O$^+$, respectively, where we have summed-over the fine-structure $J$ of the
$^4\rmD$ term, for convenience of comparison. The DR is taken from the partial {\it adf09} files of  \citet{Zat04a},
archived by \citet{Badn06a}. As in the case of totals, we see that the partial DR
is enhanced below $\sim 1000$~K by fine-structure transitions. The ($LS$-coupling) $R$-matrix results of \citet{Nah99}
lie 15\% and 35\% below our RR results for the ground and excited state, respectively, at temperatures below $\sim 1000$~K. 

Recombination into the $2\rms^2 2\rmp^2 3\rmp ^4\rmD_J$ states of O$^+$ is of interest because of the discrepancy in 
temperatures needed to model recombination line and forbidden line abundances in the Ring Nebula NGC 6720 \citep{Gar01}. 
The relevant recombination line (emissivity) is for the subsequent  $3\rmp ^4\rmD$ to $3\rms ^4\rmP$ transition at 
$\lambda 4661$ in O{\sc ii}. The $\lambda 4661$ emissivity used by \citet{Gar01} is based-on the effective recombination rate
coefficient of \citet{Sto94}, whose calculations were carried-out in $LS$-coupling. We note (not shown) only a $\approx 10\%$ enhancement
of the zero-density non-cascade partial sum (RR+DR) at $10^4$~K, on going from $LS$- to intermediate-coupling. This enhancement
becomes a factor of two down at $10^3$~K, but it is probably too low a temperature to help resolve the discrepancy in
the Ring Nebula. Interestingly, our partial sum (DR+RR) rate coefficient to $2\rms^22\rmp^23\rmp ^4\rmD$ agrees to about 10\%
with the $\lambda 4661$ effective recombination rate coefficient of \citet{Sto94}, over the temperature range of
$5\times 10^3 - 2\times 10^4$~K for which he tabulates results. Since density effects are negligible for
this line and the branching ratio for $3\rmp ^4\rmD$ to $3\rms ^4\rmP$ is approximately unity, we infer that the cascade
correction is not large. Work on a detailed revised $\lambda 4661$ effective recombination rate coefficient is in progress \citep{Sto06}.

\section{Concluding Remarks}
We have outlined the database of  RR rate coefficients
which we have established using {\sc autostructure}. Although we have highlighted some differences
from previous results which we have observed, the bulk of the groundstate total data is in good accord
with the extant. It is hoped that the availability of comprehensive partial final-state resolved and metastable totals will
stimulate the advanced modelling of non-equilibrium astrophysical plasmas, viz., dynamic, finite-density. 
Together with complementary data for DR, 
new ionization balances are already being determined for both (electron) collisional \citep{Bry06} and 
photoionized \citep{Fer06} plasmas.

\acknowledgements

I would like to thank Randall Smith and Daniel P\'{e}quignot for their interest in the partial 
and metastable RR rate coefficients, respectively. I would also like to thank Keith Butler for
supplying his $R$-matrix photoionization cross sections of O in numerical form.
This work was supported in part by PPARC Grant No. PPA$\backslash$G$\backslash$S2003$\backslash$00055
with the University of Strathclyde.

\vfill
\eject

\clearpage

\clearpage
\LongTables
\tabletypesize{\tiny}
\begin{deluxetable}{rrllllll}
\tablecaption{
Fit coefficients for total groundstate RR rate coefficients --- see Equations (\ref{fit1}) and (\ref{fit2}).
$Z$ denotes the nuclear charge and $N$ the number of electrons on the ion before recombination.
\label{tab1}
}
\tablewidth{0pt}
\tablehead{
  \colhead{$Z$} & \colhead{$N$} &    \colhead{$A$}     &   \colhead{$B$}    &    \colhead{$T_0$}
     &    \colhead{$T_1$}     &   \colhead{$C$}    &    \colhead{$T_2$}      }
\startdata
  1 & 0 & 8.318(-11)\tablenotemark{a} & 0.7472 & 2.965(0) & 7.001(5) &        &            \\
  2 & 0 & 1.818(-10) & 0.7492 & 1.017(1) & 2.786(6) &        &            \\
  3 & 0 & 2.867(-10) & 0.7493 & 2.108(1) & 6.268(6) &        &            \\
  4 & 0 & 3.375(-10) & 0.7475 & 4.628(1) & 1.121(7) &        &            \\
  5 & 0 & 4.647(-10) & 0.7484 & 6.142(1) & 1.753(7) &        &            \\
  6 & 0 & 5.337(-10) & 0.7485 & 9.502(1) & 2.517(7) &        &            \\
  7 & 0 & 6.170(-10) & 0.7481 & 1.316(2) & 3.427(7) &        &            \\
  8 & 0 & 6.552(-10) & 0.7470 & 1.951(2) & 4.483(7) &        &            \\
  9 & 0 & 8.218(-10) & 0.7491 & 2.046(2) & 5.638(7) &        &            \\
 10 & 0 & 8.278(-10) & 0.7470 & 2.991(2) & 7.006(7) &        &            \\
 11 & 0 & 9.743(-10) & 0.7488 & 3.217(2) & 8.428(7) &        &            \\
 12 & 0 & 1.022(-09) & 0.7476 & 4.098(2) & 1.011(8) &        &            \\
 13 & 0 & 1.134(-09) & 0.7482 & 4.619(2) & 1.179(8) &        &            \\
 14 & 0 & 1.261(-09) & 0.7488 & 5.068(2) & 1.365(8) &        &            \\
 15 & 0 & 1.269(-09) & 0.7472 & 6.495(2) & 1.581(8) &        &            \\
 16 & 0 & 1.432(-09) & 0.7485 & 6.688(2) & 1.793(8) &        &            \\
 17 & 0 & 1.487(-09) & 0.7489 & 7.847(2) & 2.012(8) &        &            \\
 18 & 0 & 1.615(-09) & 0.7483 & 8.433(2) & 2.266(8) &        &            \\
 19 & 0 & 1.623(-09) & 0.7473 & 1.024(3) & 2.543(8) &        &            \\
 20 & 0 & 1.809(-09) & 0.7490 & 1.025(3) & 2.788(8) &        &            \\
 21 & 0 & 1.752(-09) & 0.7470 & 1.303(3) & 3.095(8) &        &            \\
 22 & 0 & 1.899(-09) & 0.7476 & 1.348(3) & 3.393(8) &        &            \\
 23 & 0 & 2.051(-09) & 0.7490 & 1.389(3) & 3.691(8) &        &            \\
 24 & 0 & 2.046(-09) & 0.7476 & 1.638(3) & 4.027(8) &        &            \\
 25 & 0 & 2.306(-09) & 0.7499 & 1.546(3) & 4.333(8) &        &            \\
 26 & 0 & 2.275(-09) & 0.7481 & 1.836(3) & 4.736(8) &        &            \\
 27 & 0 & 2.234(-09) & 0.7461 & 2.190(3) & 5.152(8) &        &            \\
 28 & 0 & 2.577(-09) & 0.7497 & 1.949(3) & 5.434(8) &        &            \\
 29 & 0 & 2.456(-09) & 0.7477 & 2.414(3) & 5.896(8) &        &            \\
 30 & 0 & 2.854(-09) & 0.7491 & 2.125(3) & 6.271(8) &        &            \\
 36 & 0 & 3.140(-09) & 0.7485 & 3.537(3) & 9.062(8) &        &            \\
 42 & 0 & 3.583(-09) & 0.7474 & 5.008(3) & 1.235(9) &        &            \\
 54 & 0 & 5.120(-09) & 0.7493 & 6.912(3) & 2.036(9) &        &            \\
  2 & 1 & 5.235(-11) & 0.6988 & 7.301(0) & 4.475(6) & 0.0829 & 1.682(5)  \\
  3 & 1 & 9.349(-11) & 0.6916 & 3.470(1) & 7.329(6) & 0.0351 & 6.293(5)  \\
  4 & 1 & 1.324(-10) & 0.6806 & 8.525(1) & 1.510(7) &        &            \\
  5 & 1 & 1.729(-10) & 0.6776 & 1.552(2) & 2.065(7) &        &            \\
  6 & 1 & 2.044(-10) & 0.6742 & 2.647(2) & 2.773(7) &        &            \\
  7 & 1 & 2.388(-10) & 0.6732 & 3.960(2) & 3.583(7) &        &            \\
  8 & 1 & 2.652(-10) & 0.6705 & 5.842(2) & 4.559(7) &        &            \\
  9 & 1 & 3.128(-10) & 0.6712 & 7.227(2) & 5.587(7) &        &            \\
 10 & 1 & 3.415(-10) & 0.6706 & 9.552(2) & 6.778(7) &        &            \\
 11 & 1 & 3.879(-10) & 0.6718 & 1.133(3) & 8.008(7) &        &            \\
 12 & 1 & 4.214(-10) & 0.6713 & 1.396(3) & 9.433(7) &        &            \\
 13 & 1 & 4.578(-10) & 0.6707 & 1.673(3) & 1.096(8) &        &            \\
 14 & 1 & 4.870(-10) & 0.6697 & 2.026(3) & 1.265(8) &        &            \\
 15 & 1 & 5.520(-10) & 0.6724 & 2.146(3) & 1.423(8) &        &            \\
 16 & 1 & 5.546(-10) & 0.6692 & 2.754(3) & 1.633(8) &        &            \\
 17 & 1 & 6.110(-10) & 0.6712 & 2.956(3) & 1.810(8) &        &            \\
 18 & 1 & 6.405(-10) & 0.6698 & 3.425(3) & 2.037(8) &        &            \\
 19 & 1 & 7.022(-10) & 0.6721 & 3.602(3) & 2.234(8) &        &            \\
 20 & 1 & 7.045(-10) & 0.6693 & 4.391(3) & 2.500(8) &        &            \\
 21 & 1 & 8.052(-10) & 0.6750 & 4.198(3) & 2.669(8) &        &            \\
 22 & 1 & 8.152(-10) & 0.6728 & 4.937(3) & 2.950(8) &        &            \\
 23 & 1 & 8.373(-10) & 0.6711 & 5.622(3) & 3.232(8) &        &            \\
 24 & 1 & 8.771(-10) & 0.6722 & 6.118(3) & 3.478(8) &        &            \\
 25 & 1 & 9.127(-10) & 0.6717 & 6.691(3) & 3.775(8) &        &            \\
 26 & 1 & 9.983(-10) & 0.6754 & 6.651(3) & 4.017(8) &        &            \\
 27 & 1 & 1.017(-09) & 0.6738 & 7.469(3) & 4.347(8) &        &            \\
 28 & 1 & 1.063(-09) & 0.6744 & 7.979(3) & 4.653(8) &        &            \\
 29 & 1 & 1.078(-09) & 0.6736 & 8.924(3) & 4.988(8) &        &            \\
 30 & 1 & 1.142(-09) & 0.6750 & 9.193(3) & 5.308(8) &        &            \\
 36 & 1 & 1.245(-09) & 0.6716 & 1.601(4) & 7.664(8) &        &            \\
 42 & 1 & 1.536(-09) & 0.6764 & 2.012(4) & 1.008(9) &        &            \\
 54 & 1 & 2.278(-09) & 0.6872 & 2.680(4) & 1.560(9) &        &            \\
  3 & 2 & 8.700(-12) & 0.3640 & 1.470(2) & 7.153(6) & 0.1508 & 7.154(5)  \\
  4 & 2 & 1.861(-11) & 0.4052 & 5.365(2) & 2.254(7) &        &            \\
  5 & 2 & 3.251(-11) & 0.4558 & 9.164(2) & 1.749(7) &        &            \\
  6 & 2 & 4.798(-11) & 0.4834 & 1.355(3) & 1.872(7) &        &            \\
  7 & 2 & 6.245(-11) & 0.4985 & 1.957(3) & 2.177(7) &        &            \\
  8 & 2 & 8.193(-11) & 0.5165 & 2.392(3) & 2.487(7) &        &            \\
  9 & 2 & 9.958(-11) & 0.5274 & 3.012(3) & 2.896(7) &        &            \\
 10 & 2 & 1.186(-10) & 0.5354 & 3.647(3) & 3.365(7) &        &            \\
 11 & 2 & 1.393(-10) & 0.5433 & 4.258(3) & 3.872(7) &        &            \\
 12 & 2 & 1.602(-10) & 0.5492 & 4.944(3) & 4.434(7) &        &            \\
 13 & 2 & 1.761(-10) & 0.5514 & 5.962(3) & 5.081(7) &        &            \\
 14 & 2 & 2.017(-10) & 0.5588 & 6.494(3) & 5.693(7) &        &            \\
 15 & 2 & 2.190(-10) & 0.5607 & 7.591(3) & 6.396(7) &        &            \\
 16 & 2 & 2.362(-10) & 0.5615 & 8.776(3) & 7.208(7) &        &            \\
 17 & 2 & 2.594(-10) & 0.5663 & 9.623(3) & 7.952(7) &        &            \\
 18 & 2 & 2.812(-10) & 0.5686 & 1.063(4) & 8.779(7) &        &            \\
 19 & 2 & 3.046(-10) & 0.5712 & 1.160(4) & 9.639(7) &        &            \\
 20 & 2 & 3.250(-10) & 0.5743 & 1.276(4) & 1.051(8) &        &            \\
 21 & 2 & 3.366(-10) & 0.5721 & 1.473(4) & 1.163(8) &        &            \\
 22 & 2 & 3.658(-10) & 0.5767 & 1.540(4) & 1.250(8) &        &            \\
 23 & 2 & 3.815(-10) & 0.5761 & 1.718(4) & 1.368(8) &        &            \\
 24 & 2 & 3.963(-10) & 0.5760 & 1.914(4) & 1.479(8) &        &            \\
 25 & 2 & 4.386(-10) & 0.5838 & 1.888(4) & 1.556(8) &        &            \\
 26 & 2 & 4.458(-10) & 0.5802 & 2.155(4) & 1.701(8) &        &            \\
 27 & 2 & 4.629(-10) & 0.5805 & 2.352(4) & 1.831(8) &        &            \\
 28 & 2 & 4.901(-10) & 0.5836 & 2.460(4) & 1.937(8) &        &            \\
 29 & 2 & 5.041(-10) & 0.5828 & 2.701(4) & 2.082(8) &        &            \\
 30 & 2 & 5.258(-10) & 0.5832 & 2.882(4) & 2.221(8) &        &            \\
 36 & 2 & 6.300(-10) & 0.5851 & 4.346(4) & 3.140(8) &        &            \\
 42 & 2 & 8.161(-10) & 0.5972 & 5.082(4) & 4.054(8) &        &            \\
 54 & 2 & 1.082(-09) & 0.6049 & 8.335(4) & 6.391(8) &        &            \\
  4 & 3 & 1.793(-10) & 0.7669 & 8.938(-01)& 2.138(6) & 0.0310 & 1.382(4)  \\
  5 & 3 & 3.417(-10) & 0.7512 & 3.604(0) & 2.821(6) & 0.0351 & 1.295(7)  \\
  6 & 3 & 1.120(-10) & 0.6737 & 1.115(2) & 5.938(6) &        &            \\
  7 & 3 & 1.533(-10) & 0.6682 & 1.823(2) & 7.751(6) &        &            \\
  8 & 3 & 1.724(-10) & 0.6556 & 3.372(2) & 1.030(7) &        &            \\
  9 & 3 & 2.244(-10) & 0.6574 & 4.147(2) & 1.268(7) &        &            \\
 10 & 3 & 2.755(-10) & 0.6586 & 5.102(2) & 1.535(7) &        &            \\
 11 & 3 & 3.171(-10) & 0.6576 & 6.526(2) & 1.848(7) &        &            \\
 12 & 3 & 3.445(-10) & 0.6553 & 8.693(2) & 2.196(7) &        &            \\
 13 & 3 & 3.916(-10) & 0.6565 & 1.025(3) & 2.544(7) &        &            \\
 14 & 3 & 4.633(-10) & 0.6602 & 1.088(3) & 2.896(7) &        &            \\
 15 & 3 & 4.972(-10) & 0.6581 & 1.329(3) & 3.346(7) &        &            \\
 16 & 3 & 5.511(-10) & 0.6598 & 1.492(3) & 3.755(7) &        &            \\
 17 & 3 & 6.038(-10) & 0.6611 & 1.670(3) & 4.213(7) &        &            \\
 18 & 3 & 6.557(-10) & 0.6624 & 1.868(3) & 4.693(7) &        &            \\
 19 & 3 & 7.052(-10) & 0.6613 & 2.094(3) & 5.246(7) &        &            \\
 20 & 3 & 7.848(-10) & 0.6644 & 2.173(3) & 5.749(7) &        &            \\
 21 & 3 & 8.463(-10) & 0.6653 & 2.355(3) & 6.304(7) &        &            \\
 22 & 3 & 9.040(-10) & 0.6666 & 2.565(3) & 6.864(7) &        &            \\
 23 & 3 & 9.483(-10) & 0.6655 & 2.859(3) & 7.531(7) &        &            \\
 24 & 3 & 9.718(-10) & 0.6643 & 3.285(3) & 8.221(7) &        &            \\
 25 & 3 & 1.065(-09) & 0.6668 & 3.327(3) & 8.827(7) &        &            \\
 26 & 3 & 1.186(-09) & 0.6713 & 3.253(3) & 9.392(7) &        &            \\
 27 & 3 & 1.181(-09) & 0.6681 & 3.839(3) & 1.023(8) &        &            \\
 28 & 3 & 1.259(-09) & 0.6699 & 3.991(3) & 1.094(8) &        &            \\
 29 & 3 & 1.249(-09) & 0.6668 & 4.697(3) & 1.181(8) &        &            \\
 30 & 3 & 1.381(-09) & 0.6705 & 4.543(3) & 1.249(8) &        &            \\
 36 & 3 & 1.674(-09) & 0.6723 & 6.836(3) & 1.776(8) &        &            \\
 42 & 3 & 1.926(-09) & 0.6717 & 9.982(3) & 2.408(8) &        &            \\
 54 & 3 & 2.883(-09) & 0.6818 & 1.355(4) & 3.811(8) &        &            \\
  5 & 4 & 1.998(-09) & 0.8277 & 1.269(-02) & 1.016(6) & 0.0901 & 3.058(4)  \\
  6 & 4 & 2.067(-09) & 0.8012 & 1.643(-01) & 2.172(6) & 0.0427 & 6.341(4)  \\
  7 & 4 & 7.923(-10) & 0.7768 & 3.750(0) & 3.468(6) & 0.0223 & 7.206(4)  \\
  8 & 4 & 3.955(-09) & 0.7813 & 6.821(-01) & 5.076(6) &        &            \\
  9 & 4 & 3.298(-09) & 0.7702 & 2.103(0) & 6.441(6) &        &            \\
 10 & 4 & 2.557(-09) & 0.7601 & 6.293(0) & 8.091(6) &        &            \\
 11 & 4 & 1.654(-09) & 0.7508 & 2.276(1) & 9.950(6) &        &            \\
 12 & 4 & 3.989(-10) & 1.0231 & 2.601(1) & 1.227(7) &        &            \\
 13 & 4 & 1.578(-09) & 0.7394 & 6.157(1) & 1.427(7) &        &            \\
 14 & 4 & 1.851(-09) & 0.7384 & 6.906(1) & 1.644(7) &        &            \\
 15 & 4 & 1.543(-09) & 0.7315 & 1.330(2) & 1.926(7) &        &            \\
 16 & 4 & 1.740(-09) & 0.7303 & 1.494(2) & 2.193(7) &        &            \\
 17 & 4 & 2.027(-09) & 0.7307 & 1.541(2) & 2.470(7) &        &            \\
 18 & 4 & 2.165(-09) & 0.7307 & 1.801(2) & 2.753(7) &        &            \\
 19 & 4 & 2.224(-09) & 0.7299 & 2.212(2) & 3.069(7) &        &            \\
 20 & 4 & 2.421(-09) & 0.7278 & 2.438(2) & 3.424(7) &        &            \\
 21 & 4 & 2.253(-09) & 0.7253 & 3.425(2) & 3.802(7) &        &            \\
 22 & 4 & 2.463(-09) & 0.7249 & 3.638(2) & 4.164(7) &        &            \\
 23 & 4 & 2.660(-09) & 0.7253 & 3.897(2) & 4.538(7) &        &            \\
 24 & 4 & 2.747(-09) & 0.7251 & 4.436(2) & 4.921(7) &        &            \\
 25 & 4 & 2.589(-09) & 0.7216 & 5.887(2) & 5.400(7) &        &            \\
 26 & 4 & 3.322(-09) & 0.7264 & 4.563(2) & 5.746(7) &        &            \\
 27 & 4 & 3.885(-09) & 0.7301 & 4.100(2) & 6.101(7) &        &            \\
 28 & 4 & 3.531(-09) & 0.7260 & 5.650(2) & 6.627(7) &        &            \\
 29 & 4 & 3.862(-09) & 0.7272 & 5.631(2) & 7.100(7) &        &            \\
 30 & 4 & 4.233(-09) & 0.7283 & 5.568(2) & 7.545(7) &        &            \\
 36 & 4 & 4.645(-09) & 0.7261 & 1.018(3) & 1.083(8) &        &            \\
 42 & 4 & 6.974(-09) & 0.7318 & 9.622(2) & 1.438(8) &        &            \\
 54 & 4 & 1.052(-08) & 0.7363 & 1.319(3) & 2.321(8) &        &            \\
  6 & 5 & 2.995(-09) & 0.7849 & 6.670(-03) & 1.943(6) & 0.1597 & 4.955(4)  \\
  7 & 5 & 2.410(-09) & 0.7948 & 1.231(-01) & 3.016(6) & 0.0774 & 1.016(5)  \\
  8 & 5 & 2.501(-09) & 0.7844 & 5.235(-01) & 4.470(6) & 0.0447 & 1.642(5)  \\
  9 & 5 & 2.236(-09) & 0.7725 & 1.828(0) & 5.982(6) & 0.0319 & 2.529(5)  \\
 10 & 5 & 1.127(-09) & 0.7556 & 1.311(1) & 8.047(6) & 0.0250 & 2.771(5)  \\
 11 & 5 & 4.040(-09) & 0.7665 & 2.908(0) & 1.043(7) &        &            \\
 12 & 5 & 3.859(-09) & 0.7579 & 5.587(0) & 1.235(7) &        &            \\
 13 & 5 & 2.284(-09) & 0.7480 & 2.178(1) & 1.460(7) &        &            \\
 14 & 5 & 1.688(-09) & 0.7390 & 5.549(1) & 1.716(7) &        &            \\
 15 & 5 & 1.948(-09) & 0.7375 & 6.372(1) & 1.947(7) &        &            \\
 16 & 5 & 1.702(-09) & 0.7301 & 1.138(2) & 2.253(7) &        &            \\
 17 & 5 & 1.655(-09) & 0.7262 & 1.636(2) & 2.543(7) &        &            \\
 18 & 5 & 1.780(-09) & 0.7252 & 1.933(2) & 2.837(7) &        &            \\
 19 & 5 & 2.079(-09) & 0.7245 & 1.954(2) & 3.156(7) &        &            \\
 20 & 5 & 2.019(-09) & 0.7215 & 2.644(2) & 3.506(7) &        &            \\
 21 & 5 & 2.055(-09) & 0.7200 & 3.236(2) & 3.860(7) &        &            \\
 22 & 5 & 1.959(-09) & 0.7169 & 4.381(2) & 4.258(7) &        &            \\
 23 & 5 & 2.131(-09) & 0.7166 & 4.679(2) & 4.637(7) &        &            \\
 24 & 5 & 2.302(-09) & 0.7167 & 5.002(2) & 5.030(7) &        &            \\
 25 & 5 & 2.450(-09) & 0.7160 & 5.442(2) & 5.456(7) &        &            \\
 26 & 5 & 2.199(-09) & 0.7118 & 7.810(2) & 5.946(7) &        &            \\
 27 & 5 & 2.654(-09) & 0.7148 & 6.726(2) & 6.351(7) &        &            \\
 28 & 5 & 2.660(-09) & 0.7150 & 7.865(2) & 6.761(7) &        &            \\
 29 & 5 & 2.679(-09) & 0.7128 & 9.094(2) & 7.300(7) &        &            \\
 30 & 5 & 2.916(-09) & 0.7150 & 9.113(2) & 7.726(7) &        &            \\
 36 & 5 & 3.422(-09) & 0.7122 & 1.533(3) & 1.107(8) &        &            \\
 42 & 5 & 4.461(-09) & 0.7147 & 1.874(3) & 1.481(8) &        &            \\
 54 & 5 & 5.917(-09) & 0.7162 & 3.263(3) & 2.390(8) &        &            \\
  7 & 6 & 6.387(-10) & 0.7308 & 9.467(-02) & 2.954(6) & 0.2440 & 6.739(4)  \\
  8 & 6 & 2.096(-09) & 0.7668 & 1.602(-01) & 4.377(6) & 0.1070 & 1.392(5)  \\
  9 & 6 & 1.468(-09) & 0.7652 & 1.284(0) & 6.039(6) & 0.0664 & 2.235(5)  \\
 10 & 6 & 1.861(-09) & 0.7593 & 2.504(0) & 8.037(6) & 0.0406 & 3.255(5)  \\
 11 & 6 & 1.077(-09) & 0.7469 & 1.430(1) & 1.018(7) & 0.0289 & 3.593(5)  \\
 12 & 6 & 9.133(-10) & 0.7353 & 3.674(1) & 1.263(7) & 0.0211 & 4.049(5)  \\
 13 & 6 & 4.100(-09) & 0.7507 & 5.134(0) & 1.603(7) &        &            \\
 14 & 6 & 2.100(-09) & 0.7401 & 2.523(1) & 1.842(7) &        &            \\
 15 & 6 & 1.763(-09) & 0.7310 & 5.234(1) & 2.117(7) &        &            \\
 16 & 6 & 1.518(-09) & 0.7246 & 9.845(1) & 2.390(7) &        &            \\
 17 & 6 & 1.440(-09) & 0.7187 & 1.522(2) & 2.712(7) &        &            \\
 18 & 6 & 1.464(-09) & 0.7144 & 2.040(2) & 3.025(7) &        &            \\
 19 & 6 & 1.461(-09) & 0.7108 & 2.734(2) & 3.371(7) &        &            \\
 20 & 6 & 1.532(-09) & 0.7089 & 3.305(2) & 3.713(7) &        &            \\
 21 & 6 & 1.432(-09) & 0.7048 & 4.761(2) & 4.102(7) &        &            \\
 22 & 6 & 1.458(-09) & 0.7005 & 5.875(2) & 4.526(7) &        &            \\
 23 & 6 & 1.532(-09) & 0.7008 & 6.698(2) & 4.908(7) &        &            \\
 24 & 6 & 1.567(-09) & 0.6986 & 7.957(2) & 5.340(7) &        &            \\
 25 & 6 & 1.642(-09) & 0.6982 & 8.948(2) & 5.767(7) &        &            \\
 26 & 6 & 1.659(-09) & 0.6958 & 1.061(3) & 6.253(7) &        &            \\
 27 & 6 & 1.672(-09) & 0.6935 & 1.254(3) & 6.763(7) &        &            \\
 28 & 6 & 1.744(-09) & 0.6928 & 1.385(3) & 7.245(7) &        &            \\
 29 & 6 & 2.128(-09) & 0.6988 & 1.157(3) & 7.589(7) &        &            \\
 30 & 6 & 1.805(-09) & 0.6906 & 1.794(3) & 8.293(7) &        &            \\
 36 & 6 & 2.330(-09) & 0.6908 & 2.631(3) & 1.171(8) &        &            \\
 42 & 6 & 2.791(-09) & 0.6893 & 3.801(3) & 1.583(8) &        &            \\
 54 & 6 & 3.626(-09) & 0.6904 & 6.985(3) & 2.542(8) &        &            \\
  8 & 7 & 6.622(-11) & 0.6109 & 4.136(0) & 4.214(6) & 0.4093 & 8.770(4)  \\
  9 & 7 & 5.595(-10) & 0.7083 & 1.462(0) & 6.258(6) & 0.1645 & 1.880(5)  \\
 10 & 7 & 8.321(-10) & 0.7254 & 3.332(0) & 8.696(6) & 0.0921 & 3.044(5)  \\
 11 & 7 & 1.087(-09) & 0.7284 & 6.132(0) & 1.088(7) & 0.0629 & 4.559(5)  \\
 12 & 7 & 7.515(-10) & 0.7203 & 2.582(1) & 1.355(7) & 0.0436 & 5.691(5)  \\
 13 & 7 & 7.728(-10) & 0.7152 & 4.818(1) & 1.672(7) & 0.0271 & 7.209(5)  \\
 14 & 7 & 7.532(-10) & 0.7072 & 8.860(1) & 1.997(7) & 0.0185 & 6.949(5)  \\
 15 & 7 & 1.570(-09) & 0.7190 & 4.177(1) & 2.438(7) &        &            \\
 16 & 7 & 1.137(-09) & 0.7080 & 1.099(2) & 2.745(7) &        &            \\
 17 & 7 & 1.146(-09) & 0.7033 & 1.593(2) & 3.034(7) &        &            \\
 18 & 7 & 1.022(-09) & 0.6948 & 2.747(2) & 3.391(7) &        &            \\
 19 & 7 & 9.722(-10) & 0.6898 & 4.087(2) & 3.753(7) &        &            \\
 20 & 7 & 9.575(-10) & 0.6843 & 5.634(2) & 4.150(7) &        &            \\
 21 & 7 & 1.011(-09) & 0.6827 & 6.726(2) & 4.518(7) &        &            \\
 22 & 7 & 1.043(-09) & 0.6798 & 8.226(2) & 4.947(7) &        &            \\
 23 & 7 & 1.049(-09) & 0.6763 & 1.032(3) & 5.381(7) &        &            \\
 24 & 7 & 1.053(-09) & 0.6736 & 1.278(3) & 5.857(7) &        &            \\
 25 & 7 & 1.089(-09) & 0.6713 & 1.488(3) & 6.327(7) &        &            \\
 26 & 7 & 1.135(-09) & 0.6705 & 1.691(3) & 6.809(7) &        &            \\
 27 & 7 & 1.177(-09) & 0.6687 & 1.921(3) & 7.331(7) &        &            \\
 28 & 7 & 1.172(-09) & 0.6652 & 2.320(3) & 7.926(7) &        &            \\
 29 & 7 & 1.295(-09) & 0.6672 & 2.319(3) & 8.388(7) &        &            \\
 30 & 7 & 1.250(-09) & 0.6640 & 2.899(3) & 8.992(7) &        &            \\
 36 & 7 & 1.497(-09) & 0.6596 & 4.992(3) & 1.277(8) &        &            \\
 42 & 7 & 1.748(-09) & 0.6593 & 7.627(3) & 1.705(8) &        &            \\
 54 & 7 & 2.624(-09) & 0.6641 & 1.134(4) & 2.699(8) &        &            \\
  9 & 8 & 3.769(-11) & 0.5559 & 1.091(1) & 6.413(6) & 0.4534 & 1.095(5)  \\
 10 & 8 & 1.773(-10) & 0.6434 & 9.924(0) & 8.878(6) & 0.2205 & 2.292(5)  \\
 11 & 8 & 3.192(-10) & 0.6726 & 1.640(1) & 1.263(7) & 0.1232 & 3.725(5)  \\
 12 & 8 & 4.031(-10) & 0.6803 & 3.205(1) & 1.626(7) & 0.0764 & 5.399(5)  \\
 13 & 8 & 4.438(-10) & 0.6789 & 6.204(1) & 1.940(7) & 0.0571 & 7.410(5)  \\
 14 & 8 & 4.615(-10) & 0.6753 & 1.143(2) & 2.377(7) & 0.0356 & 8.595(5)  \\
 15 & 8 & 5.508(-10) & 0.6748 & 1.501(2) & 2.772(7) & 0.0213 & 1.109(6)  \\
 16 & 8 & 8.773(-10) & 0.6853 & 1.115(2) & 3.386(7) &        &            \\
 17 & 8 & 7.522(-10) & 0.6767 & 2.213(2) & 3.678(7) &        &            \\
 18 & 8 & 6.805(-10) & 0.6667 & 3.866(2) & 4.062(7) &        &            \\
 19 & 8 & 6.633(-10) & 0.6599 & 5.735(2) & 4.447(7) &        &            \\
 20 & 8 & 6.699(-10) & 0.6546 & 7.761(2) & 4.861(7) &        &            \\
 21 & 8 & 6.581(-10) & 0.6494 & 1.072(3) & 5.292(7) &        &            \\
 22 & 8 & 6.854(-10) & 0.6464 & 1.313(3) & 5.733(7) &        &            \\
 23 & 8 & 7.140(-10) & 0.6441 & 1.579(3) & 6.195(7) &        &            \\
 24 & 8 & 6.832(-10) & 0.6367 & 2.162(3) & 6.770(7) &        &            \\
 25 & 8 & 7.236(-10) & 0.6359 & 2.448(3) & 7.264(7) &        &            \\
 26 & 8 & 7.556(-10) & 0.6351 & 2.800(3) & 7.742(7) &        &            \\
 27 & 8 & 7.582(-10) & 0.6307 & 3.402(3) & 8.406(7) &        &            \\
 28 & 8 & 7.889(-10) & 0.6292 & 3.842(3) & 8.956(7) &        &            \\
 29 & 8 & 8.107(-10) & 0.6290 & 4.369(3) & 9.503(7) &        &            \\
 30 & 8 & 8.533(-10) & 0.6278 & 4.759(3) & 1.011(8) &        &            \\
 36 & 8 & 1.033(-09) & 0.6234 & 8.342(3) & 1.417(8) &        &            \\
 42 & 8 & 1.256(-09) & 0.6232 & 1.220(4) & 1.882(8) &        &            \\
 54 & 8 & 1.688(-09) & 0.6241 & 2.245(4) & 2.986(8) &        &            \\
 10 & 9 & 1.295(-11) & 0.3556 & 6.739(1) & 7.563(6) & 0.6472 & 1.598(5)  \\
 11 & 9 & 5.176(-11) & 0.4811 & 7.751(1) & 1.351(7) & 0.3467 & 3.140(5)  \\
 12 & 9 & 1.249(-10) & 0.5600 & 7.748(1) & 2.015(7) & 0.1917 & 5.139(5)  \\
 13 & 9 & 2.011(-10) & 0.5984 & 1.013(2) & 2.635(7) & 0.1072 & 7.591(5)  \\
 14 & 9 & 2.468(-10) & 0.6113 & 1.649(2) & 3.231(7) & 0.0636 & 9.837(5)  \\
 15 & 9 & 2.981(-10) & 0.6173 & 2.372(2) & 3.807(7) & 0.0361 & 1.306(6)  \\
 16 & 9 & 3.384(-10) & 0.6175 & 3.380(2) & 4.347(7) & 0.0225 & 1.672(6)  \\
 17 & 9 & 4.354(-10) & 0.6262 & 3.604(2) & 5.251(7) &        &            \\
 18 & 9 & 4.228(-10) & 0.6181 & 5.880(2) & 5.535(7) &        &            \\
 19 & 9 & 4.044(-10) & 0.6099 & 9.347(2) & 5.922(7) &        &            \\
 20 & 9 & 3.992(-10) & 0.6025 & 1.362(3) & 6.373(7) &        &            \\
 21 & 9 & 4.070(-10) & 0.5968 & 1.819(3) & 6.864(7) &        &            \\
 22 & 9 & 4.281(-10) & 0.5954 & 2.234(3) & 7.270(7) &        &            \\
 23 & 9 & 4.227(-10) & 0.5885 & 3.007(3) & 7.863(7) &        &            \\
 24 & 9 & 4.341(-10) & 0.5846 & 3.714(3) & 8.452(7) &        &            \\
 25 & 9 & 4.710(-10) & 0.5863 & 4.081(3) & 8.874(7) &        &            \\
 26 & 9 & 4.791(-10) & 0.5823 & 4.967(3) & 9.535(7) &        &            \\
 27 & 9 & 5.045(-10) & 0.5808 & 5.617(3) & 1.015(8) &        &            \\
 28 & 9 & 5.081(-10) & 0.5766 & 6.786(3) & 1.088(8) &        &            \\
 29 & 9 & 5.402(-10) & 0.5769 & 7.375(3) & 1.150(8) &        &            \\
 30 & 9 & 5.284(-10) & 0.5705 & 9.182(3) & 1.239(8) &        &            \\
 36 & 9 & 6.842(-10) & 0.5709 & 1.489(4) & 1.680(8) &        &            \\
 42 & 9 & 8.282(-10) & 0.5708 & 2.249(4) & 2.193(8) &        &            \\
 54 & 9 & 1.109(-09) & 0.5712 & 4.285(4) & 3.443(8) &        &            \\
 11 &10 & 5.095(-12) & 0.0000 & 3.546(2) & 2.310(6) & 0.9395 & 4.297(5)  \\
 12 &10 & 1.345(-11) & 0.1074 & 7.877(2) & 7.925(7) & 0.4631 & 5.027(5)  \\
 13 &10 & 3.107(-11) & 0.2854 & 8.207(2) & 8.117(7) & 0.2631 & 7.693(5)  \\
 14 &10 & 5.134(-11) & 0.3678 & 1.009(3) & 8.514(7) & 0.1646 & 1.084(6)  \\
 15 &10 & 7.550(-11) & 0.4244 & 1.206(3) & 8.603(7) & 0.1008 & 1.498(6)  \\
 16 &10 & 9.565(-11) & 0.4517 & 1.599(3) & 9.252(7) & 0.0612 & 1.986(6)  \\
 17 &10 & 1.144(-10) & 0.4665 & 2.108(3) & 9.895(7) & 0.0362 & 2.397(6)  \\
 18 &10 & 1.313(-10) & 0.4722 & 2.767(3) & 1.066(8) & 0.0208 & 2.725(6)  \\
 19 &10 & 1.590(-10) & 0.4867 & 3.096(3) & 1.192(8) &        &            \\
 20 &10 & 1.622(-10) & 0.4804 & 4.449(3) & 1.204(8) &        &            \\
 21 &10 & 1.647(-10) & 0.4734 & 6.182(3) & 1.253(8) &        &            \\
 22 &10 & 1.730(-10) & 0.4687 & 7.898(3) & 1.302(8) &        &            \\
 23 &10 & 1.800(-10) & 0.4650 & 9.950(3) & 1.357(8) &        &            \\
 24 &10 & 1.868(-10) & 0.4603 & 1.233(4) & 1.432(8) &        &            \\
 25 &10 & 1.924(-10) & 0.4562 & 1.514(4) & 1.509(8) &        &            \\
 26 &10 & 2.034(-10) & 0.4548 & 1.751(4) & 1.579(8) &        &            \\
 27 &10 & 2.125(-10) & 0.4518 & 2.042(4) & 1.667(8) &        &            \\
 28 &10 & 2.208(-10) & 0.4504 & 2.355(4) & 1.749(8) &        &            \\
 29 &10 & 2.271(-10) & 0.4464 & 2.753(4) & 1.853(8) &        &            \\
 30 &10 & 2.397(-10) & 0.4458 & 3.040(4) & 1.941(8) &        &            \\
 36 &10 & 3.051(-10) & 0.4446 & 5.297(4) & 2.530(8) &        &            \\
 42 &10 & 3.541(-10) & 0.4350 & 8.882(4) & 3.334(8) &        &            \\
 54 &10 & 4.923(-10) & 0.4400 & 1.645(5) & 5.012(8) &        &            \\
 12 &11 & 5.452(-11) & 0.6845 & 5.637(0) & 1.551(6) & 0.3945 & 8.360(5)  \\
 13 &11 & 3.171(-11) & 0.4493 & 1.821(2) & 3.529(7) & 0.2642 & 7.190(5)  \\
 14 &11 & 6.739(-11) & 0.4931 & 2.166(2) & 4.491(7) & 0.1667 & 9.046(5)  \\
 15 &11 & 1.128(-10) & 0.5330 & 2.618(2) & 4.963(7) & 0.0994 & 1.244(6)  \\
 16 &11 & 1.588(-10) & 0.5584 & 3.350(2) & 5.188(7) & 0.0591 & 1.656(6)  \\
 17 &11 & 1.790(-10) & 0.5641 & 5.390(2) & 5.733(7) & 0.0340 & 2.038(6)  \\
 18 &11 & 2.156(-10) & 0.5708 & 7.027(2) & 6.146(7) & 0.0169 & 2.845(6)  \\
 19 &11 & 2.516(-10) & 0.5771 & 8.854(2) & 6.829(7) &        &            \\
 20 &11 & 2.527(-10) & 0.5711 & 1.364(3) & 6.926(7) &        &            \\
 21 &11 & 2.578(-10) & 0.5655 & 1.955(3) & 7.157(7) &        &            \\
 22 &11 & 2.711(-10) & 0.5628 & 2.552(3) & 7.416(7) &        &            \\
 23 &11 & 2.792(-10) & 0.5588 & 3.360(3) & 7.736(7) &        &            \\
 24 &11 & 3.003(-10) & 0.5585 & 3.987(3) & 8.062(7) &        &            \\
 25 &11 & 3.197(-10) & 0.5582 & 4.712(3) & 8.394(7) &        &            \\
 26 &11 & 3.133(-10) & 0.5507 & 6.295(3) & 9.035(7) &        &            \\
 27 &11 & 3.340(-10) & 0.5516 & 7.145(3) & 9.384(7) &        &            \\
 28 &11 & 3.626(-10) & 0.5536 & 7.769(3) & 9.785(7) &        &            \\
 29 &11 & 3.761(-10) & 0.5509 & 9.021(3) & 1.036(8) &        &            \\
 30 &11 & 3.885(-10) & 0.5492 & 1.041(4) & 1.095(8) &        &            \\
 36 &11 & 4.936(-10) & 0.5469 & 1.900(4) & 1.455(8) &        &            \\
 42 &11 & 6.295(-10) & 0.5525 & 2.751(4) & 1.846(8) &        &            \\
 54 &11 & 8.258(-10) & 0.5485 & 5.795(4) & 2.911(8) &        &            \\
\enddata
\tablenotetext{a}{Note, $x.y(n)$ denotes $x.y \times 10^n$.}
\end{deluxetable}

\end{document}